\documentclass[aps, pre, superscriptaddress, 11pt, a4paper, oneside]{article}
\usepackage{amsfonts,amsmath,amssymb,amsthm,latexsym,epsfig,cancel,epstopdf}
\usepackage{setspace}
\usepackage{pdfpages}
\usepackage{bm}
\usepackage{amsbsy}
\usepackage[mathcal]{eucal}
\usepackage{graphicx}
\usepackage{caption}
\usepackage{subcaption}
\usepackage[margin=0.7in]{geometry}
\usepackage{float}
\usepackage[english]{babel} 
\addto\captionsenglish {}
\usepackage{indentfirst}

\begin{document}

\title{Survival Probability for Open Spherical Billiards} 

\begin{singlespace}

\author{Carl P. Dettmann and Mohammed R. Rahman \\
\it{School of Mathematics, University of Bristol, University Walk, Bristol BS8 1TW, U.K.}}

\end{singlespace}

\date{}
\maketitle

\begin{abstract}

We study the survival probability for long times in an open spherical billiard, extending previous work on the circular billiard. We provide details of calculations regarding two billiard configurations, specifically a sphere with a circular hole and a sphere with a square hole. The constant terms of the long-term survival probability expansions have been derived analytically. Terms that vanish in the long time limit are investigated analytically and numerically, leading to connections with the Riemann hypothesis.

\end{abstract}

\section{Introduction} \label{paper}

A mathematical billiard is a dynamical system within which a particle is in motion via alternating straight line movements in its interior and mirror-like reflections with its boundary without losing speed \cite{S. Tabachnikov}. There are many dynamical properties that are possibly present within such systems (regular, chaotic, etc.) which are obtained depending on their shapes \cite{S. Tabachnikov}. Important applications include microwave experiments \cite{S. Bittner; B. Dietz; M. Miski-Oglu; P. Oria Iriarte; A. Richter; and F. Schäfer} and microlasers \cite{P. Aschiéri; and V. Doya}.

The circular billiard is a simple but important example of regular dynamics. Orbits in the circular billiard are related to the study of mushroom billiards, since circular orbits are present in the caps of such billiards' configuration, which are a prominent example of sharply divided phase space \cite{C.P. Dettmann; and O. Georgiou2}, and widely studied both classically and quantum mechanically \cite{V. Zharnitsky}. There, typical values of a control parameter allow the existence of marginally unstable periodic orbits (MUPOs) that exhibit stickiness, specifically that unstable orbits approach regular regions in phase space \cite{C.P. Dettmann; and O. Georgiou1}. In addition, MUPOs are present in an annular billiard \cite{Altmann et al.}, within which orbits resemble those from the circular billiard. MUPOs have been realised in the context of directional emission in dielectric microcavities \cite{C.P. Dettmann; and O. Georgiou1}. The drive-belt stadium billiard has similar properties to its straight counterpart including hyperbolicity and mixing, as well as intermittency due to MUPOs whereas the big distinction between the straight and drivebelt cases is the presence of multiple MUPO families in the drivebelt \cite{C.P. Dettmann; and O. Georgiou2}. In each of these examples, the MUPOs correspond to periodic orbits of a corresponding circular billiard.

Perturbations of the class of such closed systems by the introduction of a small hole, referred to as open systems \cite{Dettmann.arx; and Georgiou.arx}, allows us to probe their internal dynamical nature. We will denote the probability of survival for time $t$ in the circular billiard by $P_{c}(t)$. The density of orbits implies that $P_{c}(t) \to 0$ as $t \to \infty$ since unperturbed periodic orbits constitute a zero-measure set in phase space. Furthermore, the leading coefficient of $P_{c}(t)$ is related to the Riemann hypothesis \cite{L.A. Bunimovich; and C.P. Dettmann}, perhaps the greatest unsolved problem in number theory \cite{J.B. Conrey}. 

Here, we study the survival probability for the spherical billiard, showing that this is also related to the Riemann hypothesis. The spherical billiard is of particular interest for applications, e.g. whispering gallery mode emission from a spherical microcavity \cite{Y.P. Rakovich1; L. Yang1; E.M. McCabe; J.F. Donegan; T. Perova; A. Moore; N. Gaponik; and A. Rogach}, while simple enough as a starting point for the study of open three dimensional billiards. There are, however, a number of qualitative differences between two and three dimensional billiards. For example, the defocusing phenomenon for generating chaos is much more involved \cite{L.A. Bunimovich; and J. Rehacek}. In our case, we note that while most orbits in the circle are dense, no orbits are dense in the sphere.

In this paper, we consider the survival probability of a spherical billiard by reducing it to a modified circle problem. A circular hole in the spherical billiard is considered in Section \ref{Spherecirc}, while a square hole is analogously considered in Section \ref{sbsh}. Concluding remarks are provided in Section \ref{ref: conclusion}.

\section{The Spherical Billiard with a Circular Hole}\label{Spherecirc}

\subsection{Hole size in the corresponding circular billiard}

We reduce the sphere problem to a circle problem. The billiard particle always remains on the same plane, defined by the initial position (relative to the center) and velocity of the particle. 

\begin{figure}[H]
\centering
\begin{subfigure}{0.3795\textwidth}
\includegraphics[width=\textwidth]{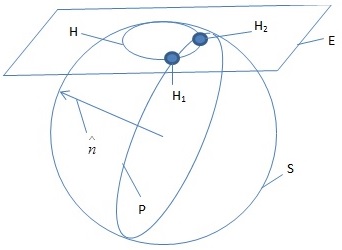}
\caption{3D view} 
\label{fig:Planeeq_a}
\end{subfigure}

\begin{subfigure}{0.3795\textwidth}
\includegraphics[width=\textwidth]{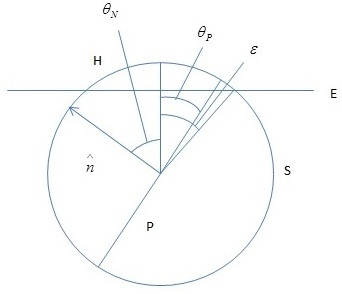}
\caption{View of sphere in $x$-$z$ plane}
\label{fig:Planeeq_b}
\end{subfigure}

\begin{subfigure}{0.3975\textwidth}
\includegraphics[width=\textwidth]{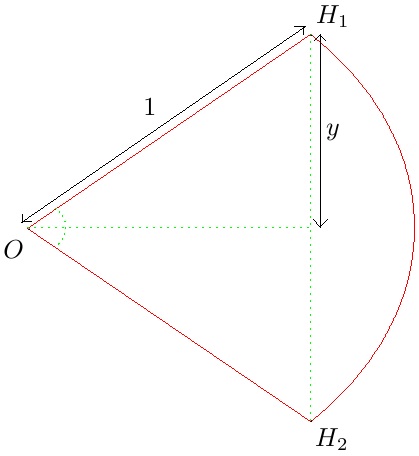}
\caption{View above sphere normal to the plane $P$}
\label{fig:Sphereintang}
\end{subfigure}
\caption{The sphere with hole of angular radius $\epsilon$ centred on the positive $z$-axis.  The particle moves in a plane $P$, which without loss of generality is
assumed to have a normal vector $\hat{n}$ in the $x$-$z$ plane}
\label{fig:Planeeq}
\end{figure}

Our construction is as follows (illustrated in figure \ref{fig:Planeeq}). We have a unit sphere, $S = \{ x,y,z | x^2+y^2+z^2=1 \}$, with a circular hole at the top of angular size $\epsilon$ (the set $H=\{ x,y,z | x^2+y^2+z^2 = 1, z \in [\cos^{-1}{(\epsilon)},1] \}$). We define the plane $E$ as $z = \cos{(\epsilon)}$, which intersects $S$ at the boundary of $H$. Due to the symmetry of our system, a particle in the sphere is confined to motion in a plane $P$, with an associated unit normal vector $\mathbf{\hat{n}} = (\tilde{x}, \tilde{y}, \tilde{z})$ (justifiable using the billiard reflection law as well as by invoking angular momentum conservation). Therefore, the equation of the plane $P$ is $\tilde{x} x+ \tilde{y} y + \tilde{z} z = 0$. In the open case, we need to consider the intersection of this plane $P$ with $H$. This intersection of $P$ with $H$ depends on the inclination of $P$ from the vertical axis at angles $\theta_{P} \in [0,\epsilon]$ (i.e. parametrized by a unit vector normal to $P$, which takes angles $\theta_{N} = \frac{\pi}{2} - \theta_{P} $ so $\theta_{N} \in [\frac{\pi}{2} - \epsilon,\frac{\pi}{2}]  $), and $P \cap S$ is also a unit circle. If $\mathbf{p}$ is a vector parallel to $P$, $\mathbf{\hat{n}} \cdot \mathbf{p} = 0$ and hence $\theta_{P} = \frac{\pi}{2} - \cos^{-1}{(\tilde{z})} $. 

Without loss of generality, we can let $\mathbf{\hat{n}} = (\tilde{x},0,\tilde{z})$ ($\tilde{x} \neq 0$ by assumption of an intersection of $P$ with $H$). The equation of the plane $P$, under the assumption that $\tilde{y} = 0$, is:

\begin{equation} \label{eq:Peq}
\tilde{x} x + \tilde{z} z = 0
\end{equation}

$ \implies x= \frac{-\tilde{z} z}{\tilde{x}}$ and furthermore on the plane $E$, $x = \frac{-\tilde{z} \cos{\epsilon}}{\tilde{x}}$. Hence, by the aid of the spherical symmetry, we obtain the $y$ coordinates of the points of intersection of the plane $P$ with $E$ and $S$:

\begin{eqnarray} \label{eq:y}
\centering
\text{(\ref{eq:Peq})} \implies y^{2} = \sin^2{(\epsilon)} - \frac{\tilde{z}^{2}}{\tilde{x}^{2}} \cos^{2}{(\epsilon)} \implies y =  \pm \sqrt{1 - \frac{\cos^2{(\epsilon)}}{\cos^2{(\theta_P)}}}
\end{eqnarray}

using the fact that $\mathbf{\hat{n}}$ is a unit vector and $\theta_{P}$ is the angle the plane $P$ makes with the z-axis. 

If we observe the top of the sphere at a point perpendicular to the plane $P$, we find (as illustrated in figure \ref{fig:Sphereintang}) that

\begin{equation} \label{eq:H1toH2angle}
\angle H_1 O H_2 = 2 \cos^{-1} \Bigg(\frac{\cos{(\epsilon)}}{\cos{(\theta_P)}} \Bigg) 
\end{equation}

so that the coordinates of $H_1$ and $H_2$ are $(\pm \cos{(\epsilon)} \tan{(\theta_{P})}, \pm \sqrt{1 - \frac{\cos^2{(\epsilon)}}{\cos^2{(\theta_P)}}}, \cos{(\epsilon)})$ (as illustrated in figures \ref{fig:Planeeq_a}-\ref{fig:Planeeq_b} (view at a reasonable height and at the $xz$ plane view) and \ref{fig:Sphereintang}) where $H_1$ and $H_2$ are the points on $E$ at which the plane $P$ intersects $E$ and the sphere $S$ and $O$ is the origin. 

As a result of these calculations, we are confined to a circular billiard problem of hole size $h = \angle H_1 O H_2$ calculated above.

\subsection{Circular Billiard Survival Probability measure}

In this escape problem, initial conditions are distributed with respect to a specified probability measure $\mu$, so that if the set of initial conditions remaining until time $t$ is denoted $M_t$ the survival probability is given by $P(t)=\mu(M_t)$. As noted in \cite{L.A. Bunimovich; and C.P. Dettmann}, for a circle or sphere this may be weighted by any smooth function of the conserved (angular momentum) variable(s) and remain invariant. In particular, the equilibrium measure for the spherical billiard flow gives the probability $\mathbb{P}_{sph}(r<R)=R^3$, which differs from that of the circle ($\mathbb{P}_{circ}(r<R)=R^2$).  This means that a weighted measure is needed for the circular billiard arising above, which we now calculate. Let $\psi\in[0,\pi/2)$ be the angle of incidence of the particle with the sphere, that is, between the velocity and the normal at the point of collision, $r\in[0,1)$ the distance of an initial point from the center, $\alpha\in[0,\pi]$ the angle between the velocity and the radial vector (assuming $r>0$) and $\phi$ the azimuthal angle relative to the radial vector.  The distance of closest approach to the center, which is also the magnitude of the conserved angular momentum, is $\sin\psi$. The symmetry implies that the angles corresponding to the location of the particles may be integrated out, leaving 

\begin{eqnarray} \label{eq: P_sinpsi_g_s}
\centering
\mathbb{P}(\sin\psi>s) & = & \frac{\int_s^1 r^2dr \int_{\frac{\pi}{2}-\cos^{-1}{\big(\frac{s}{r}\big)}}^{\frac{\pi}{2}+\cos^{-1}{\big(\frac{s}{r}\big)}} \sin\alpha d\alpha \int_0^{2\pi} d\phi}{\int_0^1 r^2dr \int_0^\pi \sin\alpha d\alpha \int_0^{2\pi} d\phi} \nonumber \\
& = & (1-s^2)^{\frac{3}{2}}.
\end{eqnarray}

Thus we have

\begin{equation}
\mathbb{P}(\psi>\Psi)=\mathbb{P}(\sin\psi>\sin\Psi)=\cos^3 \Psi
\end{equation}

so that the probability density at each periodic orbit $\psi_{m,n}=\pi/2-m\pi/n$ is $3\cos(m\pi/n)\sin^2(m\pi/n)$.  Here, $m$ and $n$ are coprime integers so that $0<m<n/2$, since if $m=n/2$ the result is zero.  A calculation similar to \cite{L.A. Bunimovich; and C.P. Dettmann} for the circular billiard with this initial measure and hole of angular size $h$ gives 

\begin{subequations}
\begin{align}
P_{c}(t) &\sim \frac{3}{2\pi t} \sum_{n=1}^\infty \sum_{\shortstack{$m = 0$\\ $(m,n)=1$}}^{\lceil \frac{n}{2} - 1 \rceil} n G \Bigg( \frac{2\pi}{n}  - h \Bigg) \sin^{3}{\frac{\pi m}{n}} \cos{\frac{\pi m}{n}} \label{eq:P_c r a} \\
&= \frac{3}{2\pi t} \sum_{n=3}^\infty n G \Bigg( \frac{2\pi}{n}  - h \Bigg) \sum_{d | n} \mu(d) \Bigg[ \frac{1}{4} \Bigg( \frac{\sin{\frac{2 \pi}{n} } - \sin{ \frac{2 \pi}{n} \Bigg(\lfloor \frac{n}{2} \rfloor + 1 \Bigg)} + \sin{\frac{2 \pi}{n}}  \lfloor \frac{n}{2} \rfloor}{2(1 - \cos{\frac{2 \pi}{n}})} \nonumber \\
&- \frac{1}{4} \Bigg( \frac{2 \sin{\frac{4 \pi}{n} } - 2 \sin{ \frac{4 \pi}{n} } \Bigg(\lfloor \frac{n}{2} \rfloor + 1 \Bigg) + 2 \sin{\frac{4 \pi}{n} } \lfloor \frac{n}{2} \rfloor}{2(1 - \cos{\frac{4 \pi}{n}})} \Bigg) \Bigg) \Bigg] \equiv \frac{B_c}{t} \label{eq: P_c r b},
\end{align}
\end{subequations}

as $t\to\infty$, where

\begin{equation}
G(x)=\left\{\begin{array}{cc}
x^2&x>0\\
0&x<0
\end{array}\right.
\end{equation}

and $\mu$ is the M\"{o}bius function, defined by $\mu(1) = 1$, $\mu(p) = -1$ for primes $p$ and $\mu(mn) = \mu(m)\mu(n)$ if gcd($m,n$) = 1; otherwise $\mu(mn) = 0$. In addition, $\mu(n)$, is an important multiplicative function in number theory and combinatorics. The German mathematician August Ferdinand M\"{o}bius introduced it in 1832. The function has many interesting properties, including it being expressible as a sum of exponentials without directly knowing the factorization of its argument \cite{Hardy G. H.; and Wright E. M.}. In addition, the derivation of equation (\ref{eq: P_c r b}) is provided via equations (\ref{eq: sin3cos}) - (\ref{eq: g_cases}) in Appendix \ref{app:CBSPC}.

We ask the following question: How does $P(t)$ behave as $h \to 0$? We expect $P_c(t) \sim \frac{C}{h t}$ from \cite{L.A. Bunimovich; and C.P. Dettmann}. 

Noting that large integers are coprime with asymptotic probability $\frac{6}{\pi^2}$ \cite{C. Pickover} and that these large values dominate at small $h$, we can replace the sums to leading order by integrals obtaining an approximation given by: 

\begin{eqnarray} \label{eq:P_c a}
B_c  \approx B^a_{c} & = & \frac{3}{2 \pi} \int_{0}^{\frac{2 \pi}{h}} d n \int_{0}^{\frac{1}{2}} n \, ds \, n \Bigg(\frac{2 \pi}{n} - h \Bigg)^2 \sin^{3}{(\pi s)} \cos{(\pi s)} \frac{6}{\pi^2} \label{eq: P_c a init} \nonumber \\
& = & \frac{6}{\pi h} 
\end{eqnarray}

where $s = \frac{m}{n}$. From the above measure, a lower weight of density of initial conditions is near the center of the circle. By a Mellin (essential tool in probability theory \cite{Galambos J; Simonelli I}) and M\"{o}bius transform approach we can derive a more precise asymptotic expansion for $t P^a_c (t)$ in the limit of $h \to 0$ (equation (\ref{eq: res}) from Appendix \ref{app:CBSPC}). We thereby find the following asymptotic form in the limit of $h \to 0$ (where $\zeta$ denotes the Riemann Zeta function \cite{G Beliakov; Y Matiyasevich, R. V. Ramos, J.B. Conrey, L.A. Bunimovich; and C.P. Dettmann}):

\begin{equation} \label{eq: P_adv}
P(h,t) \approx P^m_c(t) = \frac{1}{t} \Bigg(\frac{6}{\pi h} + \sum_{T > 0: \zeta{(\frac{1}{2} + i T)} = 0} A_T \cos{(B_T - T \ln{(h)})} h^{\frac{1}{2}} + \frac{\pi h \ln{(h)}}{4} + C h + D h^2 \Bigg). 
\end{equation}

\begin{figure}[H]
\centering
\includegraphics[width=0.50\textwidth]{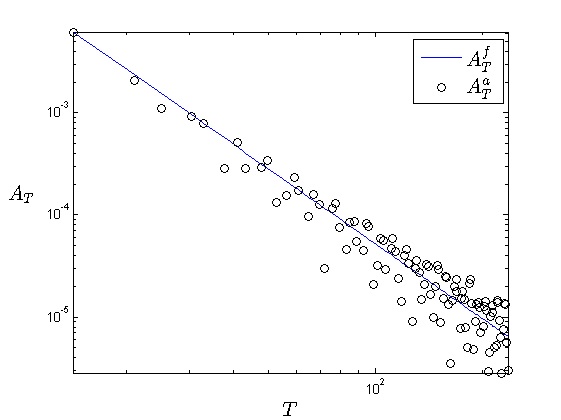}
\caption{Variation of $A_T$ with $T$}
\label{fig:ATvT}
\end{figure}

We make particular use of zeros of $\zeta{(s)}$, specifically the trivial zeros at the negative even integers \cite{G Beliakov; Y Matiyasevich} and the non-trivial zeros with real part 1/2 assuming the Riemann hypothesis \cite{J.B. Conrey}. In equation (\ref{eq: P_adv}), $A_T$ and $B_T$ arise from the residue calculations involving the non-trivial zeros of $\zeta{(s)}$ provided in Appendix \ref{app:CBSPC} \cite{R. V. Ramos}. The measure of contribution from the first several non-trivial zeros of $\zeta{(s)}$ to $P(h,t)$ is provided in table \ref{table: nontriv} (where $\zeta{(\frac{1}{2}+iT)} = 0$ and $T > 0$) \cite{Odlyzko}. A plot of how the amplitude of the first 100 of these contributions, $A_T$, vary with $T$ is represented in figure \ref{fig:ATvT}, where $A^a_T$ represents our analytic approximation to $A_T$ and $A^f_T$ represents a linear fit to $\log{(A_T)}$ with respect to $\log{(T)}$ over the first 10 positive imaginary parts of the non-trivial zeros of $\zeta{(s)}$. The fit obtained and used is $A^f_T = 0.0062 \Big(\frac{14.1347}{T} \Big)^{2.437}$. 

\begin{table}[ht] 
\caption{Coefficients in equation (\ref{eq: P_adv}).} 
\centering 
\begin{tabular}{| c | c | c |} 
\hline 
$T$ & $A_T$ & $B_T$ \\ [0.5ex] 
\hline
14.135 & 0.617466 $\times 10^{-2}$ & 2.0965 \\
21.022 & 0.205564 $\times 10^{-2}$ & -2.1446 \\
25.011 & 0.11114 $\times 10^{-2}$ & 1.4902 \\
30.425 & 0.92912 $\times 10^{-3}$ & 2.0580 \\
32.935 & 0.77994 $\times 10^{-3}$ & -1.6065 \\
37.586 & 0.28235 $\times 10^{-3}$ & -1.7963 \\
40.919 & 0.51293 $\times 10^{-3}$ & 1.1931 \\
43.327 & 0.28679 $\times 10^{-3}$ & -2.5958 \\
49.774 & 0.33897 $\times 10^{-3}$ &  -0.79929 \\
52.970 & 0.13183 $\times 10^{-3}$ & 1.8771 \\
56.446 & 0.15387 $\times 10^{-3}$ & 0.78320 \\ 
59.347 & 0.23170 $\times 10^{-3}$ & 3.1401 \\ 
60.832 & 0.17208 $\times 10^{-3}$ & -1.0177 \\ 
65.113 & 0.96093 $\times 10^{-4}$ & -2.0131 \\ 
67.080 & 0.15754 $\times 10^{-3}$ & 0.18616 \\ 
69.546 & 0.12575 $\times 10^{-3}$ & 2.3221 \\ 
72.067 & 0.29663 $\times 10^{-4}$ & -1.1540 \\ 
75.705 & 0.11696 $\times 10^{-3}$ & -2.9898 \\ 
77.145 & 0.12989 $\times 10^{-3}$ & -0.98383 \\ 
79.337 & 0.75163 $\times 10^{-4}$ & 1.0051 \\
\hline 
\end{tabular}
\label{table: nontriv} 
\end{table}

\begin{figure}[H]
\centering
\includegraphics[width=0.50\textwidth]{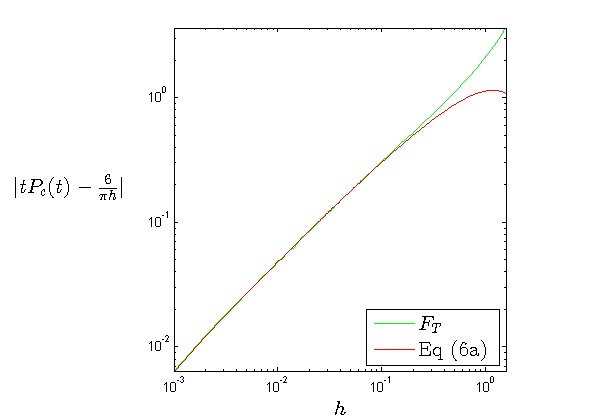}
\caption{Superimposed plots of the small $h$ asymptotics of $\big|t P_c(t) - \frac{6}{\pi h} \big|$ vs $h$, where the version for equation (\ref{eq:P_c r a}) is represented.}
\label{fig:P_c v h}
\end{figure}

We plot $\big|t P_c(t) - \frac{6}{\pi h} \big|$ from equation (\ref{eq:P_c r a}) and using $F_T$ ($P^m_c(t)$, with fitted $C$ and $D$, and taking zeros $|T| \leq 236.52$). In addition, we have

\begin{equation} \label{eq: FT}
F_T = \Bigg|\sum_{T > 0: \zeta{(\frac{1}{2} + i T)} = 0}^{\infty} A_T \cos{(B_T - T \ln{(h)})} h^{\frac{1}{2}} + \frac{\pi h \ln{(h)}}{4} - 1.158 h-0.9594 h^2 \Bigg|.
\end{equation}

It is found that $F_T$ converges to equation (\ref{eq:P_c r a}) for smaller values of $h$, as we expect. These fits are plotted in figure \ref{fig:P_c v h}. Note that equations (\ref{eq: P_adv}) and (\ref{eq: FT}) involve the factor $h^{\frac{1}{2}}$, which assumes the truth of the Riemann hypothesis, the assertion that all nontrivial zeros of $\zeta{(s)}$ have $Re{(s)} = \frac{1}{2}$. The presence of a zero with $Re{(s)} = z > \frac{1}{2}$ would lead to a term with $h^{1 - z}$.

We will now use the results regarding the survival probability in order to obtain results for our non-trivial extension, the spherical billiard survival probability.

\subsection{Extension from Circular to Spherical Billiard Survival Probability}

We will denote the survival probability for the spherical billiard with a circular hole for time $t$ by $P_{sc}(t)$. The limit of $P_{sc}(t)$ as $t \to \infty$ is:

\begin{eqnarray} \label{eq: Psclim}
P_{sc}(t) & \sim & \frac{\int_{0}^{2 \pi} \int_{\epsilon}^{\frac{\pi}{2}} \cos{(\theta_P)} \,d\theta_P \,d\phi}{\int_{0}^{2 \pi} \int_{0}^{\frac{\pi}{2}} \cos{(\theta_P)} \,d\theta_P \,d\phi} \nonumber \\
& = & 1 - \sin{(\epsilon)}
\end{eqnarray}

The integration over $\theta_P$ in equation (\ref{eq: Psclim}) is carried out since an initial condition of a trajectory in terms of its position and velocity is uniformly distributed within the sphere and lies on a unit circular plane (by the billiard reflection law). Unlike the circular billiard, we see that $1 - \sin{(\epsilon)}$ is the fraction of initial conditions that never escapes from the spherical billiard. We will now seek a more detailed long time behaviour for $P_{sc}(t)$.

We will assume that $\epsilon \ll 1$. To aid our analysis, we utilize the given \cite{L.A. Bunimovich; and C.P. Dettmann} following asymptotic expression for the long time survival probability of a trajectory in the circular billiard with a hole of size $h \ll 1$ such that $h t \gg 1$:

\begin{equation} \label{eq:Pcu}
 P^{u}_{c}(t) \sim \left\{
  \begin{array}{ll}
    \frac{C}{h t} + O \Big(\frac{1}{t^2}\Big) & \mbox{if \(h > 0 \)} \\
    1 & \mbox{if \(h = 0, \)} \\
  \end{array} \right.
\end{equation}

where $C$ is a constant (in the case of \cite{L.A. Bunimovich; and C.P. Dettmann}, $C = 2$ and from equation (\ref{eq:Pcu}), $C = \frac{6}{\pi}$) and a key point to stress is that the above holds if $h \gg \frac{C}{t}$.  In addition, open integrable billiards including the circle are well known to exhibit power law decay at long times. The justification of the survival probability behaving as $O(\frac{1}{t})$, in terms of integrating over allowed phase space is available \cite{L.A. Bunimovich; and C.P. Dettmann}. 

We present two approaches to estimating the survival probability $P_{sc}(t)$. The "unrefined" version, $P^u_{sc}(t)$,  assumes equation (\ref{eq:Pcu}) is valid for all time $t$. The "refined" version, $P^r_{sc}(t)$, instead assumes that $P_{sc}(t)$ is given by the minimum of $\frac{C}{ht}$ and 1 for values of $h$ that depend on a circular plane's orientation within the sphere.

Therefore, the long time survival probability for the spherical billiard is:

\begin{eqnarray}
P^u_{sc}(t) & \sim & \frac{\int_{0}^{2 \pi} \int_{\epsilon}^{\frac{\pi}{2}} \cos{(\theta_P)} \,d\theta_P \,d\phi + \int_{0}^{2 \pi} \int_{0}^{\epsilon} \frac{C}{h t} \cos{(\theta_P)} \,d\theta_P \,d\phi }{\int_{0}^{2 \pi} \int_{0}^{\frac{\pi}{2}} \cos{(\theta_P)} \,d\theta_P \,d\phi} \nonumber \\
& = & 1 - \sin{(\epsilon)}+\frac{I^u_{sc}}{t}
\end{eqnarray}

where 

\begin{eqnarray} \label{eq: B_sc_epsilon}
I^u_{sc} & = & \int_{0}^{\epsilon} \frac{C}{h} \cos{(\theta_P)} \,d\theta_P.
\end{eqnarray}

Furthermore,

\begin{equation} \label{eq:Pcr}
 P^{r}_{c}(t) \sim \left\{
  \begin{array}{ll}
    \frac{C}{h t} & \mbox{if \(ht > C \)} \\
    1 & \mbox{if \(ht < C \)}. \\
  \end{array} \right.
\end{equation}

We will show that $P^r_{sc}(t)$ and $P^u_{sc}(t)$ are equivalent up to $O \Big(\frac{1}{t^2}\Big)$ (equation (\ref{eq:err_ord})), where we use $C = \frac{6}{\pi}$ as in equation (\ref{eq:Pcu}). 

We can also obtain a numerical approximation of this $O \Big(\frac{1}{t^2}\Big)$ coefficient for all $\epsilon$ by substituting in $\angle H_1 O H_2$ into equation (4) from \cite{L.A. Bunimovich; and C.P. Dettmann}, truncating the summation in equation (4) from \cite{L.A. Bunimovich; and C.P. Dettmann} to the upper limit of $\lfloor \frac{2 \pi}{\angle H_1 O H_2} \rfloor$, multiplying this by $\cos{(\angle H_1 O H_2)}$ and use approximations via maple etc. 

We show in Appendix \ref{app: Ref_v_unr} that the refined and unrefined versions of our survival probability differ in magnitude by an amount asymptotic to $\frac{27 \cos^{2}{(\epsilon)}}{2 \pi t^{2}\sin{(\epsilon)}}$.

From the analyses in Appendix \ref{app: Ref_v_unr} we find that for large time $t$, using the full $h$-dependence of $P_c(t)$ from equation (\ref{eq: P_adv}),

\begin{equation}
P_{sc}(t) \approx 1 - \sin{(\epsilon)} + \frac{B(\epsilon)}{t} - \frac{27 \cos^{2}{(\epsilon)}}{2 \pi t^{2}\sin{(\epsilon)}},
\end{equation}

where

\begin{eqnarray} \label{eq: B_fin}
B(\epsilon) & = & \epsilon \int_0^1 \Bigg(\frac{6}{\pi h} + \sum_{\shortstack{$T > 0$\\ $\zeta{(\frac{1}{2} + i T)} = 0$}}^{T_{max}} A_T \cos{(B_T - T \ln{h})} h^{\frac{1}{2}} + \frac{\pi}{4} h \ln{h} + P_{c,h} h \Bigg) \cos{\epsilon \tau} d \tau \nonumber \\
& \approx & \epsilon \int_0^1 \Bigg(\frac{6}{\pi (2 \cos^{-1} (\frac{\cos{\epsilon}}{\cos{\epsilon \tau}}))} + \sum_{\shortstack{$T > 0$\\ $\zeta{(\frac{1}{2} + i T)} = 0$}}^{T_{max}} A_T \cos{(B_T - T \ln{(2 \epsilon \sqrt{1 - \tau^2})})} (2 \epsilon \sqrt{1 - \tau^2})^{\frac{1}{2}} \nonumber \\
& + & \frac{\pi}{4} (2 \epsilon \sqrt{1 - \tau^2}) \ln{(2 \epsilon \sqrt{1 - \tau^2})} + P_{c,h} (2 \epsilon \sqrt{1 - \tau^2}) \Bigg) d \tau \nonumber \\
& = & \frac{3}{2} + 2^{\frac{1}{2}} \int_0^1 \sum_{\shortstack{$T > 0$\\ $\zeta{(\frac{1}{2} + i T)} = 0$}}^{T_{max}} A_T \cos{(B_T - T \ln{(2 \epsilon \sqrt{1-\tau^2})})} (1 - \tau^2)^{\frac{1}{4}} \epsilon^{\frac{3}{2}} d \tau \nonumber \\
& + & \Bigg(\frac{\pi^2}{8} \ln{(2 \epsilon)} + \frac{\pi}{2} \int_0^1 \sqrt{1-\tau^2} \ln{\sqrt{1-\tau^2}} d \tau + \frac{P_{c,h} \pi}{2} - \frac{1}{2} \Bigg) \epsilon^2, \nonumber \\
\end{eqnarray} 

where $\int_0^1 (1 - \tau^2)^{\frac{1}{4}} d \tau = \frac{\sqrt{\pi} \Gamma{(\frac{1}{4})}}{6 \Gamma{(\frac{3}{4})}}$ $(\approx 0.874019)$, $T_{max} \approx 236.52$, we assume $\Big(2 \cos^{-1}{(\frac{\cos{(\epsilon)}}{\cos{(\epsilon \tau)}})} \Big) \approx 2 \epsilon \sqrt{1 - \tau^2}$; $\int_0^1 \sqrt{1 - \tau^2} d \tau = \frac{\pi}{4}$, 

\begin{equation}
\int_0^1 \sqrt{1-\tau^2} \ln{\sqrt{1-\tau^2}} d \tau \approx -0.1516974409;
\end{equation}

and $P_{c,h}=-1.158$ (the fitted $O(h)$ term's coefficient in $P_{c} (t)$ in equation (\ref{eq: FT})). We will now justify that $\Big(2 \cos^{-1}{(\frac{\cos{(\epsilon)}}{\cos{(\epsilon \tau)}})} \Big) \approx 2 \epsilon \sqrt{1 - \tau^2}$ is a good approximation for $\epsilon \ll 1$. Firstly, we find from expanding for small $\epsilon$ that $2 \cos^{-1}{\Big(\frac{\cos{(\epsilon)}}{\cos{(\epsilon \tau)}}\Big)} = 2 \epsilon \sqrt{1 - \tau^2} + \frac{\epsilon^3 \tau^2}{6} \sqrt{1 - \tau^2} + \ldots$.

\begin{figure}[H]
\centering 
\includegraphics[width=0.50\textwidth]{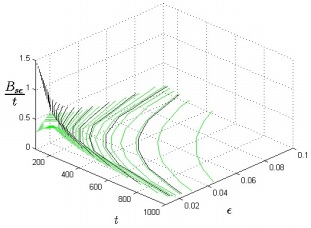}\caption{Plot of both the unrefined (darker curve) and refined (lighter curve) versions of the integral $\int_{0}^{\epsilon} \frac{C}{2 \cos^{-1} (\frac{\cos{(\epsilon)}}{\cos{(\theta_P)}}) t} \cos{(\theta_P)} \,d\theta_P$, $\frac{B_{sc}}{t}$, vs time $t$ and hole size $\epsilon$} \label{fig:Bscvteps}
\end{figure}

One can obtain a plot of both the unrefined and refined versions of the integral \\ $\int_{0}^{\epsilon} \frac{C}{2 \cos^{-1} (\frac{\cos{(\epsilon)}}{\cos{(\theta_P)}}) t} \cos{(\theta_P)} \,d\theta_P$, which contributes to the second order term of the survival probability. We will present this for various hole sizes $\epsilon$ in figure \ref{fig:Bscvteps}, where $I_{sc}(t)^u = \int_0^{\epsilon} \frac{C \cos{(\theta)}}{2 t \cos^{-1}({\frac{\cos{(\epsilon)}}{\cos{(\theta)}}})} d \theta$ and $I_{sc}(t)^r = \int_0^{g(\frac{C}{t}\epsilon)} \frac{C \cos{(\theta)}}{2 t \cos^{-1}({\frac{\cos{(\epsilon)}}{\cos{(\theta)}}})} d \theta$.

From figure \ref{fig:Bscvteps}, the unrefined version of the numeric survival probability curve seems to lie above its refined counterpart, in particular for early times, which is expected since we have long-time approximations which are not necessarily valid for short values of time. Furthermore, the unrefined version of the numeric survival probability curve seems to increase infinitely for times $t$ tending to 0, which is consistent since this version is defined as a probability for certain values of $h$ relative to $t$.

\subsection{Direct Numerics compared with Analytical Results}

\begin{figure}[H]
\centering
\includegraphics[width=0.50\textwidth]{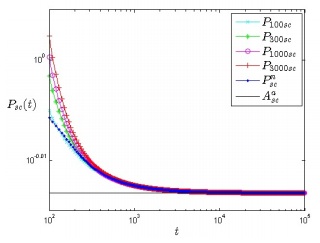}
\caption{Plot of $P_{sc}(t)$ vs $t$ for $\epsilon = 0.03$, $10^8$ numeric simulation samples, and fitting time ranges of $T$ to $10^5, T \in \{10^2, 3 \cdot 10^2, 10^3, 3 \cdot 10^3\}$}
\label{fig:spherePsuprefa}
\end{figure}

We can obtain plots of fits of $P_{sc}(t)$ from direct numerical simulations. Numerical simulations were carried out using C++. The survival times for a sample of $10^8$ initial conditions, uniformly random positions and velocities from inside the unit sphere, were plotted in cumulative distribution plots ($P^n_{sc}$). In addition, numeric survival probability limits are provided as horizontal lines ($A^n_{100sc}, A^n_{300sc}$, $A^n_{1000sc}$ and $A^n_{3000sc}$). We show this in the form of a logarithm scale plot in figure \ref{fig:spherePsuprefa} for the cases of 3 term fits over the time ranges $[T, 10^5], T \in \{100, 300, 1000, 3000 \}$ $ (P^{(2)}_{Tsc})$, where the fitting function applied to the set of all survival fractions translated by subtraction of the latest surviving fraction is of the following form:

\begin{equation} \label{eq: sphere_B_C_fitanz}
B \Bigg(\frac{1}{t} - \frac{1}{10^k} \Bigg) + C \Bigg(\frac{1}{t^2} - \frac{1}{10^{2k}} \Bigg).
\end{equation}

where $k \in \mathbb{N}$ and in this case $k = 5$.

From the above four fitting cases, figure \ref{fig:spherePsuprefa} seems to indicate that a fit of the form 

\begin{equation} \label{eq: sphfitanz}
A+\frac{B}{t} + \frac{C}{t^2}
\end{equation}

is a good approximation to the survival probability in the billiard configuration in question. One may expect better consistency from a higher order fit i.e. $\sum_{i = 0}^{M}{\frac{P_i}{t^i} }, M \geq 3$.

We will now compare fitted 2nd order coefficients of the long-time expansion of $P_{sc}(t)$ ($B_{sc}$) obtained through various fitting time ranges.

\begin{figure}[H]
\centering
\includegraphics[width=0.50\textwidth]{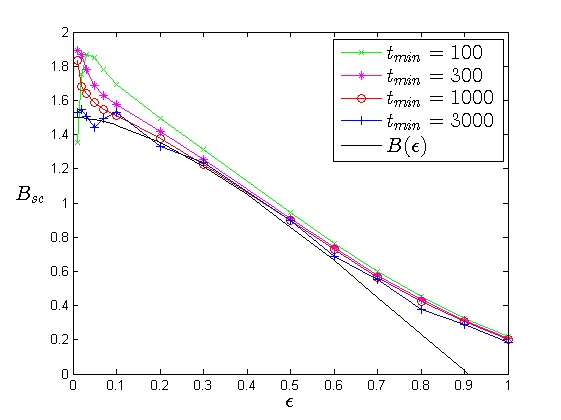}
\caption{$B_{sc}$ vs $\epsilon$ , $10^8$ samples, fits of the form $[T, 10^4]$, $T \in \{100, 300, 1000, 3000 \}$} 
\label{fig:plotsupBvepsilon}
\end{figure}

From figure \ref{fig:plotsupBvepsilon}, it is observed that the fits with smaller $t_{min}$ yield smaller estimates for $B_{sc}$, in particular for smaller spherical billiard circular hole sizes. In addition, larger discrepancies between $B(\epsilon)$ and the fits are situated at very small $\epsilon$ and $\epsilon$ close to $\frac{\pi}{2}$. Potential explanations for this include the ideas of not having an infinitely high number of initial conditions; not being able to select a fitting time range having infinitely large initial as well as final time values and the use of a finite number of terms in equation (\ref{eq: B_fin}). In addition, we find good agreement that $B(\epsilon)$ has a dependence on $\epsilon$ of the form $const_1 + const_2 \epsilon^{\frac{3}{2}}$ due to the Riemann hypothesis.

We note that for exactly $\epsilon = 0$ there is no $B$ ($O \Big(\frac{1}{t} \Big)$) coefficient. This is due to the non-commutativity in the $t \to \infty$ and $\epsilon \to 0$ limits. A similar phenomenon occurs in the stadium billiard \cite{Dettmann.arx; and Georgiou.arx}.

\begin{figure}[H]
\centering
\includegraphics[width=0.50\textwidth]{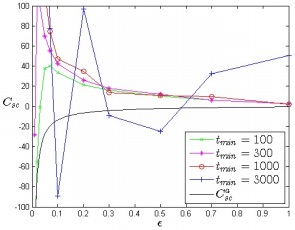}
\caption{$C_{sc} \sim \frac{const.}{\epsilon}$ (from equation (\ref{eq: C_sc})) vs $\epsilon$, $10^8$ samples, fits of the form given in equation (\ref{eq: sphere_B_C_fitanz})}
\label{fig:plotsupCvepsilon}
\end{figure}

We can also investigate the trend of the third-order term, $C_{sc}$ in the expansion of the survival probability. Figure \ref{fig:plotsupCvepsilon} shows the fitted $O(\frac{1}{t^2})$ contribution (ansatz of the form in equation (\ref{eq: sphere_B_C_fitanz})) vs $\epsilon$ curve from fitting over the same time ranges as in figure \ref{fig:plotsupBvepsilon} with $10^8$ numerical simulation samples in comparison with our theoretical coefficient for $C_{sc}, -\frac{27 \cos^{2}{(\epsilon)}}{\sin{(\epsilon)}}$ ($C^a_{sc}$).

From figure \ref{fig:plotsupCvepsilon}, we find that there must be other second order effects not taken into account. In addition, there is indication that $C_{sc} \to -\infty$ as $\epsilon \to 0$.

We have derived analytic expressions regarding the long-time, small hole-size survival probability for the spherical billiard, which indicate that the fraction of initial conditions surviving for long-time decreases approximately linearly with hole size ($1 - \sin{(\epsilon)} \approx 1 - \epsilon$) and that the rate of decay of the probability decreases approximately quadratically (the $\epsilon^2$ terms present). We will now obtain analogous results for a modified configuration. 

\section{Spherical Billiard with a Square-Shaped hole} \label{sbsh}

We will now present results based on a more non-trivial spherical billiard problem. We now let a square hole be placed at the top of the sphere. 

The equation of points on the boundary of the sphere is \(x^{2} + y^{2} + z^{2} = 1\), so \(z = \pm \sqrt{1 - x^{2} - y^{2}}\). The square-shaped hole of this billiard is mathematically defined as \(\{x, y, z : |x| < \epsilon', |y| < \epsilon', \\
z = + \sqrt{1 - x^{2} - y^{2}} \} \) or as \( \{ \theta, \phi : | \sin(\phi)\sin(\theta) | < \epsilon', | \cos(\phi)\sin(\theta) | < \epsilon', \theta \in [0,\frac{\pi}{2}) \}\).

A particle is confined to a circular plane during its motion inside the sphere. Therefore, we are reduced to a circular billiard problem. A particle can escape the billiard through its hole if the plane that it is confined to intersects the spherical billiard's hole. We let $\bf{\underline{\text{n}}}_{pvi}$ be a vector on a plane intersecting the square hole as well as pointing towards and/or through the hole, towards the topmost part of the plane. In spherical polar coordinates, a point on the surface of the sphere has position given by $(\sin{(\theta)} \cos{(\phi)},\sin{(\theta)} \sin{(\phi)},\cos{(\theta)})$ (considering a unit sphere), where $\phi$ is the angle a point in spherical polar coordinate space, makes anti-clockwise with respect to the horizontal $x$ axis and $\theta$ is the angle that the same point makes with respect to the positive vertical $z$ axis. For each $\phi$, the range of $\theta$ that is allowed to be taken by $\bf{\underline{\text{n}}}_{pvi}$ is:

\[   \theta \in \left\{   \begin{array}{ll}
    (0,\sin^{-1}({\frac{\epsilon'}{\cos{(\phi)}}})) & \mbox{ if $\phi \in [\frac{- \pi}{4}, \frac{\pi}{4}] \cup [\frac{3 \pi}{4}, \frac{5 \pi}{4}]$} \\
    (0,\sin^{-1}({\frac{\epsilon'}{\sin{(\phi)}}})) & \mbox{ if $\phi \in [\frac{- 5 \pi}{4}, \frac{-\pi}{4}] \cup [\frac{ \pi}{4}, \frac{3 \pi}{4}]$},  \end{array} \right. \]

without loss of generality.

Therefore, the range of $\theta$ that is allowed to be taken by a vector $\bf{\underline{\text{n}}}_{\perp}$ normal to the plane intersecting the hole is:

\[   \theta \in \left\{   \begin{array}{ll}
    (\frac{\pi}{2},\sin^{-1}({\frac{\epsilon'}{\cos{(\phi)}}})+\frac{\pi}{2}) & \mbox{ if $\phi \in [\frac{- \pi}{4}, \frac{\pi}{4}] \cup [\frac{3 \pi}{4}, \frac{5 \pi}{4}]$} \\
    (\frac{\pi}{2},\sin^{-1}({\frac{\epsilon'}{\sin{(\phi)}}})+\frac{\pi}{2}) & \mbox{ if $\phi \in [\frac{- 5 \pi}{4}, \frac{-\pi}{4}] \cup [\frac{ \pi}{4}, \frac{3 \pi}{4}]$}.  \end{array} \right. \]

In Cartesian coordinates we write \(\underline{\text{n}}_{\perp} = (n_{px},n_{py},n_{pz}) \) with $n_{pz} = \cos{(\theta)}$. Therefore, a vector on this plane, $(x,y,z)$ satisfies $n_{px}x + n_{py}y + n_{pz}z= 0$. 

If there is an intersection between a plane and the hole at $x = \pm \epsilon'$, 

\begin{eqnarray*}
& & n_{px}(\pm \epsilon') + n_{py}y + n_{pz}z = 0 \\
\\
& \implies & z = \frac{-n_{px} n_{pz} \epsilon' \pm \sqrt{n_{px}^{2}n_{pz}^{2} \epsilon'^{2} - (n_{py}^{2} + n_{pz}^{2})(n_{px}^{2}\epsilon'^{2}+n_{py}^{2}\epsilon'^{2} - n_{py}^{2})}}{n_{py}^{2} + n_{pz}^{2}}
\end{eqnarray*}

using $y = \pm \sqrt{1 - \epsilon'^2 - z^2}$.

Similar expressions can be made for other components of positions of intersections as well as those with the other boundaries of the square hole.

We can construct bounds on the constant term in the survival probability. The spherical-billiard-square-hole problem is one intermediate between those of circular holes of sizes $\sin^{-1}{(\epsilon')}$ and $\sin^{-1}{(\sqrt{2}\epsilon')}$. Since the measure of initial conditions that survive in the billiard decreases with increasing hole size, the bound on the constant term in the expansion of the survival probability in the spherical-billiard-square-hole problem for large time $t$ is:

\begin{equation*}
\centering
A_{ss} \in \Bigg[ \frac{\int_{0}^{2 \pi} \int_{\sin^{-1}{\sqrt{2}\epsilon'}}^{\frac{\pi}{2}} \cos{(\theta)} d\theta d\phi}{\int_{0}^{2 \pi} \int_{0}^{\frac{\pi}{2}} \cos{(\theta)} d\theta d\phi} , \frac{\int_{0}^{2 \pi} \int_{\epsilon'}^{\frac{\pi}{2}} \cos{(\theta)} d\theta d\phi}{\int_{0}^{2 \pi} \int_{0}^{\frac{\pi}{2}} \cos{(\theta)} d\theta d\phi} \Bigg] = [1 - \sqrt{2} \epsilon', 1 - \epsilon']
\end{equation*}

An asymptotic expression for the survival probability can be derived as follows:

We first consider the subregion $S_{-\frac{\pi}{4}..\frac{\pi}{4}} = \{(x,y,z) | 0<x<y, x^{2}+y^{2}+z^{2} = 1 \}$ of the sphere. The constant term is derived as follows: 

For $\tilde{\phi} \in [-\frac{\pi}{4}, 0]$, the ellipse image found by a bird's eye view grazes the square hole at $x = -y = \pm \epsilon'$.

\begin{eqnarray}
\implies A_{ss} & = & \frac{8}{2 \pi} \int_{\frac{-\pi}{4}}^{0} \int_{\tan^{-1}{\Bigg(-\sqrt{\frac{1-2\epsilon'^{2}}{\epsilon'^{2}(\cos{(\tilde{\phi})} - \sin{(\tilde{\phi})})^{2}}}\Bigg)}}^{\pi} \sin{(\tilde{\theta})} d \tilde{\theta} d \tilde{\phi} \nonumber \\
                             & = & \frac{2}{\pi} \sin^{-1}{(1 - 2 \epsilon'^2)}
\end{eqnarray}

Therefore, the survival probability in this case is 

\[
P_{ss}(t) \sim \frac{2}{\pi}\sin^{-1}{(1 - 2 \epsilon'^2)} \text{ as } t \to \infty . 
\]

\begin{figure}[H]
\centering
\includegraphics[width=0.50\textwidth]{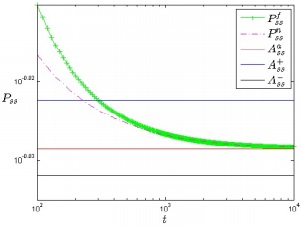}
\caption{Superimposed upper and lower bounds ($A^+_{ss}$ and $A^-_{ss}$ respectively), limit ($A_{ss}^a$) and numeric simulation ($P_{ss}^f$) of long-time survival probability, $P_{ss}$, for the spherical billiard with a square hole vs time, $t$, maximum $10^{6}$ collisions, $10^{5}$ samples and hole size $\epsilon'=0.05$}
\label{fig:plotnumeps0point05spheresquaremaxcol10tothe6samsize10tothe5trangeto10tothe4fb10t3e10t4}
\end{figure}

Numeric simulations for the spherical billiard with a square hole are presented in figure \ref{fig:plotnumeps0point05spheresquaremaxcol10tothe6samsize10tothe5trangeto10tothe4fb10t3e10t4}.

In figure \ref{fig:plotnumeps0point05spheresquaremaxcol10tothe6samsize10tothe5trangeto10tothe4fb10t3e10t4} the numerical versions of $P_{ss}$ ($P_{ss}^n$ and $P_{ss}^f$) seem to almost lie on top of one another and converge to their expected long-time limits, which falls between the expected bounds.

\begin{figure}[H]
	\centering
		\includegraphics[width=0.50\textwidth]{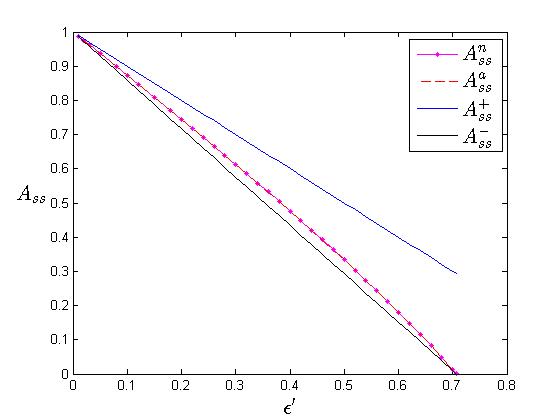}
	\caption{Analytic ($A_{ss}^a$), Numeric ($A_{ss}^n$ obtained over the time range $[10^3, 10^4]$) ((last surviving fraction of initial conditions of a large time range)-$\frac{\text{Numeric \(O(\frac{1}{t}) \) term}}{10^4}$), and upper and lower bounds ($A^+_{ss}$ and $A^-_{ss}$) of the constant term of the survival probability, $A_{ss}$, versus hole size, $\epsilon'$}
	\label{fig:plotspheresquaresupAvseps_1000}
\end{figure}

\begin{figure}[H]
\centering
\includegraphics[width=0.50\textwidth]{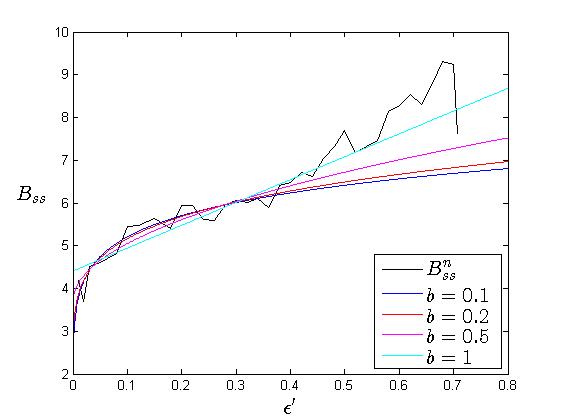}
\caption{Numeric \(O(\frac{1}{t}) \) term, $B_{ss}$ of the survival probability versus hole size, $\epsilon'$, fitting the curve $B(1/t-1/10^4)+C(1/t^2-1/10^8)$ to a survival dataset over a time range of 1000 to 10000. Also, fits of the form $B_{ss} = B_0 + a\epsilon'^b$ over the range $[0.01,\epsilon'_{max} = 0.4]$}
\label{fig:plotspheresquarenumBvseps_1000_point4}
\end{figure}

Plots related to the constant and \(O(\frac{1}{t}) \) terms of the survival probability are presented in figures \ref{fig:plotspheresquaresupAvseps_1000}-\ref{fig:plotspheresquarenumBvseps_1000_point4}.

From figure \ref{fig:plotspheresquaresupAvseps_1000}, we can see that the analytic and numeric counterparts of the constant term in the survival probability seem to more or less lie on top of each other as well as fall within the bounds as expected and therefore these superimposed plots appear consistent. Due to the lack of symmetry of the configuration in question, it has not been possible to analytically compute the \(O(\frac{1}{t}) \) term of the survival probability. The plot in figure \ref{fig:plotspheresquarenumBvseps_1000_point4} provides an indication that the \(O(\frac{1}{t}) \) term of the survival probability increases with hole size $\epsilon'$, in contrast to the circular hole case. One possible explanation of this result in the square hole is the increasing fraction of initial conditions that belong to circle billiards with small non-zero hole sizes, contributed from the sharpening of corners (due to the spherical geometry, the square has acute-angled corners, which decrease with $\epsilon'$) of square holes and hence larger areas above corners partially covered by a hole. According to the obtained fits, as the selected value of $b$ decreases, the consistency between the obtained fit and simulation data improves in the limit of  $\epsilon^{\prime} \to 0$. Furthermore, the value for the power of $\epsilon'$ (according to the fits) in the dependence of $B^n_{ss}$ on $\epsilon'$ appears to lie in the range [0.1,0.5], which is much smaller than $\frac{3}{2}$ for the case of the circular hole.

\section{Conclusion} \label{ref: conclusion}

In this work, we have investigated the survival probability for large time as well as for small hole size in the spherical billiard under the fundamental assumption of an absence in resistive forces and on the basis of derived supplementary circular billiard calculations. We have used various tools to investigate the asymptotic trend of the survival probability in the circular billiard corresponding to the plane of motion of the particle in the sphere with effective hole size $h$ which indicates that 

\begin{equation}
\centering
P(h,t) \sim \frac{6}{\pi h t}.
\end{equation}

We have found that the approximate long-time ($t \to \infty$) survival probability for the spherical billiard with a circular hole of size $\epsilon$ (i.e. $\theta \in [0, \epsilon)$, where $\theta$ is the angular distance from the north pole in spherical polar coordinates) is: 

\begin{equation*}
\centering
P_{sc}(t) \sim 1 - \sin{(\epsilon)} + \frac{B(\epsilon)}{t}
\end{equation*}

For the spherical billiard with a circular hole, the constant term of its long-time survival probability expansion decreases approximately linearly with hole size. In addition, the $O(\frac{1}{t})$ term decreases with hole size. It is found that the term $B(\epsilon)$ is dominated by the Riemann hypothesis, in that $B(\epsilon) = \frac{3}{2} + O(\epsilon^{\frac{3}{2}-\delta})$ (where $\delta > 0$ is due to the multiplicities of each of the non-trivial zeros of the Riemann Zeta function being possibly greater than one, i.e. $\zeta{(s)} = O((s-(\frac{1}{2}+iT))^k)$, $k \in \mathbb{N}$, $k > 1$).

Furthermore, analogous results have been found for a billiard configuration with a modified geometry, specifically a sphere with a square hole.

For the spherical billiard with a square hole, the constant term of its long-time survival probability expansion also decreases approximately linearly with hole size. In addition, the $O(\frac{1}{t})$ term apparently increases with hole size and at a slower rate than the case of the circular hole configuration (in accordance with numerical findings).

This work leads in a number of interesting directions. The study of cylindrical billiard (with a hole of particular shape at a particular location on its boundary) dynamics, can be studied using cylindrical polar coordinates in deriving analogous survival probability measures for the integration of horizontal circular billiard problems.

The problem of deriving the prevalence as well as the importance of the regular regions in phase space of a physical system comprising particles that are predominantly chaotic \cite{C.P. Dettmann} has been stated. In this context, there exist billiards that comprise more than one particle. For example, the distribution of incident angles of collisions between the particles and the boundary in the case of two identical particles of varying radius confined to a unit circle has been considered \cite{S. Lansel; M.A. Porter and L.A. Bunimovich}. It would be interesting to consider the corresponding three dimensional problem of two particles in a sphere. 

\section*{Acknowledgements}

We thank Orestis Georgiou for discussions and computer simulations, as well as Thomas Bloom; Yahaya Ibrahim; Chris Joyner; and Shirali Kadyrov for discussions.

\appendix

\section{Mellin transform calculations} \label{app:CBSPC} 

Let $n = p^{r_1}_1 \cdots p^{r_q}_q$, where $n, q, p_1, \ldots, p_q, r_1, \ldots, r_q \in \mathbb{N}$. Assume $t = p_{r_1} \ldots p_{r_t} | n$ where $t, p_{r_1} \ldots p_{r_t}$, $r_1 \ldots r_t \in \mathbb{N}$. The number of occurrences of $h_t$ in $\sum_{d | n}^{N} \sum_{\shortstack{$ m = d $ \\ $ d | m $}}^N \mu(d) h_m$ is $\binom{r_t}{0} + \ldots + \binom{r_t}{r_t}(-1)^{r_t} = (1-1)^{r_t} = 0$, where $\binom{r_t}{i}$ denotes terms generated by $d$ such that $d$ is a product of $i$ primes each occurring in the factorisation of $d$ once. Therefore,
\\
\\
\begin{eqnarray} \label{eq: coprime sum}
\sum_{\shortstack{$ m = 1 $ \\ $(m,n) = 1$}}^{N} h_m = \sum_{d | n}^{N} \sum_{\shortstack{$ m = d $ \\ $ d | m $}}^N \mu(d) h_m = \sum_{d | n}^N \sum_{\shortstack{$ jd = d $ \\ $ d | jd $}}^{jd = N} \mu(d) h_{jd} = \sum_{d | n}^N \sum_{ j = 1 }^{j = \lfloor \frac{N}{d} \rfloor} \mu(d) h_{jd}
\end{eqnarray}
\\
\\
\begin{eqnarray} \label{eq: sin3cos}
\sin^{3}{(\frac{\pi m}{n})} \cos{(\frac{\pi m}{n})} & = & \frac{i}{16} \Bigg(2 \Bigg(\exp{\Bigg(\frac{-i 2 \pi m}{n} \Bigg)} - \exp{\Bigg(\frac{i 2 \pi m}{n} \Bigg)} \Bigg) \nonumber \\
& + & \exp{\Bigg(\frac{i 4 \pi m}{n} \Bigg)} - \exp{\Bigg(\frac{- i 4 \pi m}{n} \Bigg)} \Bigg)
\end{eqnarray}

\begin{eqnarray}  \label{eq: gn}
g \bigg(\frac{n}{d} \bigg) & = & \sum_{m = 1}^{\lfloor \frac{n}{2d} \rfloor} \sin^{3}{(\frac{\pi m d}{n})} \cos{(\frac{\pi m d}{n})} \nonumber \\
& = & \frac{-i}{8} \Bigg(\exp{\frac{i 2 \pi d}{n}} \frac{1 - (\exp{\frac{i 2 \pi d}{n}})^{\lfloor \frac{n}{2 d} \rfloor}}{1 - \exp{\frac{i 2 \pi d}{n}}} - \exp{\frac{- i 2 \pi d}{n}} \frac{1 - (\exp{\frac{- i 2 \pi d}{n}})^{\lfloor \frac{n}{2 d} \rfloor  }}{1 - \exp{\frac{- i 2 \pi d}{n}}} \nonumber \\
& - & \frac{1}{2} \Bigg(\exp{\frac{i 4 \pi d}{n}} \frac{1 - (\exp{\frac{i 4 \pi d}{n}})^{\lfloor \frac{n}{2 d} \rfloor }}{1 - \exp{\frac{i 4 \pi d}{n}}} - \exp{\frac{- i 4 \pi d}{n}} \frac{1 - (\exp{\frac{-i 4 \pi d}{n}})^{\lfloor{ \frac{n}{2 d} \rfloor}}}{1 - \exp{\frac{-i 4 \pi d}{n}}} \Bigg) \Bigg) \nonumber \\
& = & \frac{1}{8} \Bigg( \frac{\sin{\Big(\frac{2 \pi d}{n} \Big)} - \sin{\Bigg( \Big( \frac{2 \pi d}{n} \Big) \Big(\lfloor \frac{n}{2 d} \rfloor + 1 \Big) \Bigg)} + \sin{\Bigg( \Big(\frac{2 \pi d}{n} \Big) \lfloor \frac{n}{2 d} \rfloor \Bigg)}}{(1 - \cos{\Big( \frac{2 \pi d}{n} \Big)})} \nonumber \\
& - & \frac{1}{2} \Bigg( \frac{ \sin{\Big( \frac{4 \pi d}{n} \Big)} - \sin{\Bigg( \Big( \frac{4 \pi d}{n} \Big) \Big(\lfloor \frac{n}{2 d} \rfloor + 1 \Big) \Bigg)} + \sin{\Bigg( \Big(\frac{4 \pi d}{n} \Big) \lfloor \frac{n}{2 d} \rfloor \Bigg)}}{(1 - \cos{\Big( \frac{4 \pi d}{n} \Big) })} \Bigg) \Bigg)
\end{eqnarray}
\\
\\
where we express $\sin^{3}{(\frac{\pi m}{n})} \cos{(\frac{\pi m}{n})}$ in terms of exponentials and sum the geometric series. 
\\
\\
From (\ref{eq: coprime sum}) - (\ref{eq: gn}),

\begin{equation} \label{eq: g_cases}
g \bigg(\frac{n}{d} \bigg) = \begin{cases} \frac{1}{4} \Bigg(\frac{\sin{\Big(\frac{2 \pi d}{n} \Big)}}{(1 - \cos{\Big( \frac{2 \pi d}{n} \Big) })} \Bigg)
&\mbox{ if } \frac{n}{d} \text{ is even } \\
\frac{1}{8} \Bigg(\frac{\sin{\Big(\frac{2 \pi d}{n} \Big)} +2 \sin{\Big(\frac{\pi d}{n} \Big)}}{(1 - \cos{\Big( \frac{2 \pi d}{n} \Big) })} -  \frac{\sin{\Big(\frac{4 \pi d}{n} \Big)} - 2 \sin{\Big(\frac{2 \pi d}{n} \Big)}}{2(1 - \cos{\Big( \frac{4 \pi d}{n} \Big) })} \Bigg)
&\mbox{ if } \frac{n}{d} \text{ is odd }
\end{cases}
\end{equation}

\begin{eqnarray} \label{eq: transform}
& & \sum_{\shortstack{$ m = 1 $ \\ $(m,n) = 1$}}^{\lfloor \frac{n}{2} \rfloor} \sin^{3}{(\frac{\pi m}{n})} \cos{(\frac{\pi m}{n})} \\
& = & \sum_{\shortstack{$d | n$ \\ $ \lfloor \frac{n}{2 d} \rfloor = \frac{n}{2 d} $}}^{\lfloor \frac{n}{2} \rfloor} \mu(d) \frac{1}{4} \Bigg(\frac{\sin{\Big(\frac{2 \pi d}{n} \Big)}}{(1 - \cos{\Big( \frac{2 \pi d}{n} \Big) })} \Bigg) + \sum_{\shortstack{$d | n$ \\ $ \lfloor \frac{n}{2 d} \rfloor = \frac{n}{2 d} - \frac{1}{2}$}}^{\lfloor \frac{n}{2} \rfloor} \mu(d) \frac{1}{8} \Bigg(\frac{\sin{\Big(\frac{2 \pi d}{n} \Big)} +2 \sin{\Big(\frac{\pi d}{n} \Big)}}{(1 - \cos{\Big( \frac{2 \pi d}{n} \Big) })} \nonumber  \\
& - & \frac{\sin{\Big(\frac{4 \pi d}{n} \Big)} - 2 \sin{\Big(\frac{2 \pi d}{n} \Big)}}{2 (1 - \cos{\Big( \frac{4 \pi d}{n} \Big) })} \Bigg) \nonumber \\
& = & \sum_{\shortstack{$d | n$ \\ $ \lfloor \frac{n}{2 d} \rfloor = \frac{n}{2 d} $}}^{\lfloor \frac{n}{2} \rfloor} \mu(d) \frac{1}{4} (\frac{n}{\pi d}-\frac{1}{3}\frac{\pi d}{n}-\frac{1}{45} (\frac{\pi d}{n})^3-\frac{2}{945} (\frac{\pi d}{n})^5-\frac{1}{4725} (\frac{\pi d}{n})^7 - \frac{2}{93555} (\frac{\pi d}{n})^9+O((\frac{\pi d}{n})^{10})) \nonumber \\
& + & \sum_{\shortstack{$d | n$ \\ $ \lfloor \frac{n}{2 d} \rfloor = \frac{n}{2 d} - \frac{1}{2}$}}^{\lfloor \frac{n}{2} \rfloor} \mu(d) \frac{1}{8} (2 \frac{n}{\pi d}+\frac{1}{3} \frac{\pi d}{n}+\frac{59}{360} (\frac{\pi d}{n})^3+\frac{1007}{15120} (\frac{\pi d}{n})^5 + \frac{16319}{604800} (\frac{\pi d}{n})^7+O((\frac{\pi d}{n})^8))
\nonumber \\
\nonumber \\
\end{eqnarray}
\\
\\
We can verify that $\sum_{\shortstack{$ m = 0$ \\ $(m,n) = 1$}}^{\lfloor \frac{n}{2} \rfloor} \sin^{3}{(\frac{\pi m}{n})} \cos{(\frac{\pi m}{n})}$ satisfies equation (\ref{eq: transform}) for $n = 1$ and $n = 2$. 

The relevant associated Mellin transform (as a means of expressing our sum in equation (\ref{eq: P_c r b}) as $M^{-1}(M(h))$ to extract its small $h$ expansion) is

\begin{eqnarray}
\bar{P}^g(s) & = & \int_{0}^{\infty} \lim_{t \to \infty} t P^g_c (t) h^{s-1} d h \nonumber \\
& = & \int_{0}^{\infty} \frac{3}{2\pi} \sum_{\shortstack{$n=1$\\ $(m,n)=1$\\ $m<n/2$}}^\infty nG(\frac{2\pi}{n}-h) h^{s-1} \sin^3\frac{\pi m}{n}
\cos\frac{\pi m}{n} d h \nonumber \\
& = & \int_{0}^{2\pi/n} \frac{3}{2\pi} \sum_{\shortstack{$n=1$\\ $(m,n)=1$\\ $m<n/2$}}^\infty n (\frac{2\pi}{n}-h)^2 h^{s-1} \sin^3\frac{\pi m}{n} \cos\frac{\pi m}{n} d h, \text{ } (\text{since $G(\cdot)$ is non-zero for $h \in (0, \frac{2\pi}{n})$}) \nonumber \\
& = & \sum_{\shortstack{$n=1$\\ $(m,n)=1$\\ $m<n/2$}}^\infty \frac{6 (\frac{2 \pi}{n})^{s+1}}{s(s+1)(s+2)} \sin^3\frac{\pi m}{n} \cos\frac{\pi m}{n}, \text{ } (\text{we use integration by parts by integrating $h^{s-1}$} \nonumber \\
& & \text{and $h^s$ as well as differentiating $(\frac{2\pi}{n}-h)^2$ and $(\frac{2\pi}{n}-h)$}) \nonumber \\
& = & \sum_{n=1}^\infty \frac{6 (2 \pi)^{s+1}}{n^{s+1} s(s+1)(s+2)} \Bigg( \sum_{\shortstack{$d | n$ \\ $ \lfloor \frac{n}{2 d} \rfloor = \frac{n}{2 d} $}}^{\lfloor \frac{n}{2} \rfloor} \mu(d) \frac{1}{4} \Bigg(\frac{\sin{\Big(\frac{2 \pi d}{n} \Big)}}{(1 - \cos{\Big( \frac{2 \pi d}{n} \Big) })} \Bigg) \nonumber \\
& + & \sum_{\shortstack{$d | n$ \\ $ \lfloor \frac{n}{2 d} \rfloor = \frac{n}{2 d} - \frac{1}{2}$}}^{\lfloor \frac{n}{2} \rfloor} \mu(d) \frac{1}{8} \Bigg(\frac{\sin{\Big(\frac{2 \pi d}{n} \Big)} +2 \sin{\Big(\frac{\pi d}{n} \Big)}}{(1 - \cos{\Big( \frac{2 \pi d}{n} \Big) })} -  \frac{1}{2} \frac{\sin{\Big(\frac{4 \pi d}{n} \Big)} - 2 \sin{\Big(\frac{2 \pi d}{n} \Big)}}{(1 - \cos{\Big( \frac{4 \pi d}{n} \Big) })} \Bigg) \Bigg), \nonumber \\
& & \text{(we substitute the sum over $m$ by its M\"{o}bius transformed version)} \nonumber \\
& = & \sum_{n=1}^\infty \frac{6 (2 \pi)^{s+1}}{n^{s+1} s(s+1)(s+2)} \Bigg( \sum_{\shortstack{$d | n$ \\ $ \lfloor \frac{n}{2 d} \rfloor = \frac{n}{2 d} $}}^{\lfloor \frac{n}{2} \rfloor} \mu(d) \frac{1}{4} (\frac{n}{\pi d}-\frac{1}{3}\frac{\pi d}{n}-\frac{1}{45} (\frac{\pi d}{n})^3-\frac{2}{945} (\frac{\pi d}{n})^5-\frac{1}{4725} (\frac{\pi d}{n})^7 \nonumber \\
& - & \frac{2}{93555} (\frac{\pi d}{n})^9+O((\frac{\pi d}{n})^{10})) + \sum_{\shortstack{$d | n$ \\ $ \lfloor \frac{n}{2 d} \rfloor = \frac{n}{2 d} - \frac{1}{2}$}}^{\lfloor \frac{n}{2} \rfloor} \mu(d) \frac{1}{8} (2 \frac{n}{\pi d}+\frac{1}{3} \frac{\pi d}{n}+\frac{59}{360} (\frac{\pi d}{n})^3+\frac{1007}{15120} (\frac{\pi d}{n})^5 \nonumber \\
& + & \frac{16319}{604800} (\frac{\pi d}{n})^7+O((\frac{\pi d}{n})^8))  \Bigg),  \nonumber \\
& & (\text{we substitute the trigonometric function with their expansions for small $\frac{\pi d}{n}$}) \nonumber \\
& = & \sum_{d=1}^\infty \frac{6 (2 \pi)^{s+1}}{d^{s+1} s(s+1)(s+2)} \Bigg(\sum_{\shortstack{$j=2$ \\ $ \lfloor \frac{j}{2} \rfloor = \frac{j}{2} $}}^{\infty} \mu(d) \frac{1}{4} \frac{1}{j^{s+1}} (\frac{j}{\pi}-\frac{1}{3}\frac{\pi}{j}-\frac{1}{45} (\frac{\pi }{j})^3-\frac{2}{945} (\frac{\pi}{j})^5-\frac{1}{4725} (\frac{\pi}{j})^7 - \frac{2}{93555} (\frac{\pi}{j})^9 \nonumber \\ 
& + & O((\frac{\pi}{j})^{10})) + \sum_{\shortstack{$j=3$ \\ $ \lfloor \frac{j}{2} \rfloor = \frac{j  - 1}{2} $}}^{\infty} \mu(d) \frac{1}{8} \frac{1}{j^{s+1}} (2 \frac{j}{\pi}+\frac{1}{3} \frac{\pi}{j}+\frac{59}{360} (\frac{\pi}{j})^3+\frac{1007}{15120} (\frac{\pi}{j})^5 + \frac{16319}{604800} (\frac{\pi}{j})^7 \nonumber \\
& + & O((\frac{\pi}{j})^8)) \Bigg)
\end{eqnarray}

(where we apply the transformation $j = \frac{n}{d}$). Therefore we want the residues of

\begin{eqnarray}
&h^{-s}& \sum_{d=1}^\infty \frac{6 (2 \pi)^{s+1}}{d^{s+1} s(s+1)(s+2)} \Bigg(\sum_{\shortstack{$j=2$ \\ $ \lfloor \frac{j}{2} \rfloor = \frac{j}{2} $}}^{\infty} \mu(d) \frac{1}{4} \frac{1}{j^{s+1}} (\frac{j}{\pi}-\frac{1}{3}\frac{\pi}{j}-\frac{1}{45} (\frac{\pi }{j})^3-\frac{2}{945} (\frac{\pi}{j})^5-\frac{1}{4725} (\frac{\pi}{j})^7 - \frac{2}{93555} (\frac{\pi}{j})^9 \nonumber \\
& + & O((\frac{\pi}{j})^{10})) + \sum_{\shortstack{$j=3$ \\ $ \lfloor \frac{j}{2} \rfloor = \frac{j  - 1}{2} $}}^{\infty} \mu(d) \frac{1}{8} \frac{1}{j^{s+1}} (2 \frac{j}{\pi}+\frac{1}{3} \frac{\pi}{j}+\frac{59}{360} (\frac{\pi}{j})^3+\frac{1007}{15120} (\frac{\pi}{j})^5 + \frac{16319}{604800} (\frac{\pi}{j})^7 \nonumber \\
& + & O((\frac{\pi}{j})^8))  \Bigg) \nonumber \\
& = & h^{-s} \frac{1}{\zeta{(s+1)}} \frac{6 (2 \pi)^{s+1}}{s(s+1)(s+2)} \Bigg( \frac{1}{4} \Bigg(\frac{\zeta{(s)}}{2^{s} \pi}-\frac{1}{3}\frac{\zeta{(s+2)} \pi}{2^{s+2}} - \ldots \Bigg) + \frac{1}{8} \Bigg(2 \frac{\zeta{(s)} (1 - \frac{1}{2^{s}}) - \frac{1}{1^{s}}}{\pi} \nonumber \\
& + & \frac{(\zeta{(s+2)}(1 - \frac{1}{2^{s+2}}) - \frac{1}{1^{s+2}}) \pi}{3}+\ldots \Bigg) \Bigg) \\
&& \text{(where we assume an analytic continuation over $s$ such that $Re(s) < 1$)} \nonumber
\end{eqnarray}
\\
\\
which at $s = 1 $, $s = 0 $, $s = -1$ and $s = -\frac{1}{2} + i t$ (contributing to non-trivial zeros of the Riemann Zeta function) respectively are: 
\\
\\
\begin{eqnarray} \label{eq: res}
Res_{s = 1} F(s) & = & \frac{6}{\pi h}, \nonumber \\
Res_{s = 0} F(s) & = & 0, \nonumber \\
Res_{s = -1} F(s) & = & h^1  \frac{6 (2 \pi)^{-1 + 1}}{\zeta{(0)} \cdot (-1) \cdot 1} \Bigg(\frac{1}{4} \Bigg(\frac{\zeta{(-1)}}{2^{-1} \pi}-\frac{\gamma \pi}{3 \cdot 2^{-1+2}}  - \ldots \Bigg) + \frac{1}{8} \Bigg(2 \frac{(\zeta{(-1)}(1 - \frac{1}{2^{-1}}) - \frac{1}{1^{-1}})}{\pi} \nonumber \\
& + & \frac{\Bigg(\gamma \Bigg(1 - \frac{1}{2^{-1}} \Bigg) - \frac{1}{1^{-1+2}} \Bigg) \pi}{3} + \ldots \Bigg) \Bigg) + h \Bigg(\frac{1}{\zeta{(0)}} \frac{6 (2 \pi)^{-1+1}}{(-1)(1)} \Bigg(\frac{1}{4} \Bigg(-\frac{1}{3} \frac{- \ln{(h)} \pi}{2^{-1+2}} \Bigg) \nonumber \\
& + & \frac{1}{8} \Bigg(\frac{1}{3} \frac{- \ln{(h)} \pi}{2^{-1+2}} \Bigg) \Bigg) \Bigg) \nonumber \\
& = & h^1 \frac{6 (2 \pi)^{-1 + 1}}{\zeta{(0)} \cdot (-1) \cdot 1} \Bigg(\frac{1}{4} \Bigg(\frac{\zeta{(-1)}}{2^{-1} \pi}-\frac{\gamma \pi}{3 \cdot 2^{-1+2}}  - \ldots \Bigg) + \frac{1}{8} \Bigg(2 \frac{(\zeta{(-1)}(1 - \frac{1}{2^{-1}}) - \frac{1}{1^{-1}})}{\pi} \nonumber \\
& + & \frac{\Bigg(\gamma \Bigg(1 - \frac{1}{2^{-1}} \Bigg) - \frac{1}{1^{-1+2}} \Bigg) \pi}{3} + \ldots \Bigg) \Bigg) + \frac{
\pi h \ln{(h)}}{4} \text{ and } \nonumber \\
Res_{s = -\frac{1}{2} + i T} F(s) & = & \frac{6 (2 \pi)^{-\frac{1}{2} + i t + 1} h^{-(-\frac{1}{2} + i t)}}{\zeta{'(-\frac{1}{2} + i t + 1)}(-\frac{1}{2} + i t)(-\frac{1}{2} + i t + 1)(-\frac{1}{2} + i t + 2)} \Bigg(\frac{1}{4} \Bigg(\frac{\zeta{(-\frac{1}{2} + i t )}}{2^{(-\frac{1}{2} + i t )} \pi} - \ldots \Bigg) \nonumber \\
& + & \frac{1}{8} \Bigg(2\frac{\zeta{(-\frac{1}{2} + i t )} (1 - \frac{1}{2^{(-\frac{1}{2} + i t)}})}{\pi}+ \ldots \Bigg) \Bigg) \text{ assuming the Riemann hypothesis } \nonumber \\
 \nonumber \\
 \nonumber \\
\end{eqnarray} 
\\
\\
We can analyse the asymptotics as $t \to \infty$ of the residues at the non-trivial zeros of $\zeta{(s)}$. We first invoke the following:

\begin{eqnarray*}
\chi(s) & = & \frac{\pi^{s-\frac{1}{2}} \Gamma{(\frac{1}{2} - \frac{s}{2})}}{\Gamma{(\frac{s}{2})}}, \\
| \Gamma(x+iy) | &\sim& \sqrt{2 \pi} |y|^{x-\frac{1}{2}}\exp{\Bigg(-\frac{\pi |y|}{2} \Bigg)}, \\
\zeta(\frac{1}{2} + i t) & = & \sum_{n = 1}^m \frac{1}{n^{\frac{1}{2} + i t}} + \chi(\frac{1}{2} + it) \sum_{n = 1}^m \frac{1}{n^{\frac{1}{2} -  i t}} + O(t^{-\frac{1}{4}}) \text{ where $m = \lfloor \sqrt{\frac{t}{2 \pi}} \rfloor$ } \\
& & \text{$\zeta(\frac{1}{2} + i t)$ diverges since $\sum_{n \in \mathbb{N}} \frac{1}{n^p}$ diverges for $\Re{(p)} \leq 1$, hence $\zeta'(\frac{1}{2} + i t)$ diverges due to } \\
& & \text{ $ln(n)$ arising in the summand from differentiation with respect to $t$}
\end{eqnarray*}

Furthermore,

\begin{eqnarray}
| \zeta{(\frac{1}{2} - i t)} | & = & \Bigg| -2^{(\frac{1}{2} - i t)} \sin{\Bigg(\frac{\pi (\frac{1}{2} - i t)}{2} \Bigg)} \Gamma(1 - (\frac{1}{2} - i t)) \zeta{'(1-(\frac{1}{2} - i t))}  \Bigg| \\
& \gg & O \Bigg(\exp{\Big(\frac{\pi t}{2} \Big)} \exp{\Big(-\frac{\pi t}{2} \Big)} \Bigg) = O(1), \nonumber \\
\Bigg| \zeta{\Bigg(-\frac{3}{2} + i t \Bigg)} \Bigg| &=& \Bigg| 2^{(-\frac{3}{2} + i t)} \sin{\Bigg(\frac{\pi (-\frac{3}{2} + i t)}{2} \Bigg)} \Gamma \Bigg(1 - \Bigg(-\frac{3}{2} + i t \Bigg) \Bigg) \zeta{\Bigg(1 - \Bigg(\frac{-3}{2} + i t \Bigg) \Bigg)} \Bigg| \nonumber \\
&=& O\Bigg( \exp{\Bigg(\frac{\pi t}{2} \Bigg)} t^2 \exp{\Bigg(-\frac{\pi t}{2} \Bigg)}\Bigg) = O(t^2) \nonumber
\end{eqnarray}

Therefore, we expect that,

\begin{eqnarray}
\Bigg| \frac{6 (2 \pi)^{-\frac{1}{2} + i t + 1} h^{-(-\frac{1}{2} + i t)}}{(-\frac{1}{2} + i t)(-\frac{1}{2} + i t + 1)(-\frac{1}{2} + i t + 2)} \Bigg(\frac{1}{4} \Bigg(\frac{\zeta{(-\frac{1}{2} + i t - 1)}}{2^{(-\frac{1}{2} + i t - 1)} \pi} - \ldots \Bigg) & + & \frac{1}{8} \Bigg(2\frac{\zeta{(-\frac{1}{2} + i t - 1)} (1 - \frac{1}{2^{(-\frac{1}{2} + i t - 1)}})}{\pi} \nonumber \\
& + & \ldots \Bigg) \Bigg) \Bigg| = o(\frac{1}{t}) \text{ as $t \to \infty$} \nonumber \\
\end{eqnarray}

\section{Refined vs Unrefined versions of Spherical Billiard Survival Probability} \label{app: Ref_v_unr}

The refined approximation to our survival probability gives for the corresponding integral $I_{sc}^{r}$:

\begin{equation}
I_{sc}^{r} = \int_{0}^{g(\frac{C}{t},\epsilon)} \frac{C}{2\cos^{-1}{(\frac{\cos{(\epsilon)}}{\cos{(\theta_P)}})} t}\cos{(\theta_P)} \,d\theta_P
\end{equation}

where \[  g \Bigg(u,\epsilon \Bigg) = \left\{ 
  \begin{array}{l l}
    \cos^{-1} \Big(\frac{\cos{(\epsilon)}}{\cos{(\frac{u}{2})}} \Big) & \quad \text{if \(u \leq 2\epsilon\)} \\
    0 & \quad \text{if \(u > 2\epsilon \)} \\
  \end{array} \right.
 \]

We can interpret $g \Bigg(\frac{C}{t},\epsilon \Bigg)$ as a measure of how small a circular billiard's hole size can be, as we modify its orientation parameters, in this case $\theta_{P}$. This justification allows us to treat our "refined" version of the survival probability as a higher order expansion for $t \to \infty$, which we will discover when considering the asymptotic error between our numerical candidates $P_{sc}^{u}(t)$ and $P_{sc}^{r}(t)$.

The difference between the unrefined and refined versions of the expression is:

\begin{eqnarray} \label{eq: Ierr}
\int_0^{\epsilon} \frac{C}{2\cos^{-1} (\frac{\cos{(\epsilon)}}{\cos{(\theta_P)}}) t} \cos{(\theta_P)} \,d\theta_P & - & \Bigg(\int_0^{g(\frac{C}{t},\epsilon)} \frac{C}{2\cos^{-1} (\frac{\cos{(\epsilon)}}{\cos{(\theta_P)}}) t} \cos{(\theta_P)} \,d\theta_P +\int_{g(\frac{C}{t},\epsilon)}^{\epsilon} \cos{(\theta_P)} \,d\theta_P \Bigg) \nonumber \\
P_{sc}^{u}(t)-P_{sc}^{r}(t) & = & I_{sc}^{u}(t) - I_{sc}^{r}(t) - \int_{g(\frac{C}{t},\epsilon)}^{\epsilon} \cos{(\theta_P)} \,d\theta_P \nonumber \\
& = & \int_{g(\frac{C}{t},\epsilon)}^{\epsilon} \Bigg(\frac{C}{2\cos^{-1} (\frac{\cos{(\epsilon)}}{\cos{(\theta_P)}}) t} - 1 \Bigg) \cos{(\theta_P)} \,d\theta_P.
\end{eqnarray}

We can see that $g \big(\frac{C}{t},\epsilon \big) \to \epsilon$ as $t \to \infty$.

A series expansion approximation of the integrand $\frac{\cos{( \theta_P)}}{\cos^{-1} (\frac{\cos{(\epsilon)}}{\cos{(\theta_P)}})}$ around $\theta_P = \epsilon$ (which we are interested in using since in the limit as $t \to \infty$ yields the focus on values of $\theta_P$ in the vicinity of $\epsilon$) is:

\begin{equation} \label{eq: theta_exp}
\frac{\cos{( \theta_P)}}{\cos^{-1} (\frac{\cos{(\epsilon)}}{\cos{(\theta_P)}})} = \cos{(\epsilon)} \sqrt{\frac{\cos{(\epsilon)}}{2 \sin{(\epsilon)} (\epsilon - \theta_P)}}+O(\sqrt{\theta_P-\epsilon}), 
\end{equation}

which integrated with respect to $\theta_P$ is $O(\sqrt{\theta_P-\epsilon})$. Therefore, the integration in equation (\ref{eq: Ierr}) is integrable. Furthermore, since $g \Big(\frac{C}{t},\epsilon \Big) \to \epsilon$ as $t \to \infty$ the difference between the unrefined and refined versions tends to zero as $t \to \infty$.

We can also investigate how the error between $P_{sc}^u$ and $P_{sc}^r$ decays with time. This involves evaluating:

\begin{equation}
P_{sc}^{u}(t) - P_{sc}^{r}(t) = \int_{g(\frac{C}{t},\epsilon)}^{\epsilon} \Big(\frac{C}{2\cos^{-1} (\frac{\cos{(\epsilon)}}{\cos{(\theta_P)}}) t} - 1 \Big) \cos{(\theta_P)} \,d\theta_P = O \Big(t^{-\gamma}\Big),
\end{equation}

as $t \to \infty$, where $\gamma > 0$.

The following procedure for obtaining the leading order behaviour for the error as $t \to \infty$ has been derived:

1) Set $s=\frac{1}{t}$ so that $g = g(C s, \epsilon)$.

2) The first and second derivatives of $g(C s,\epsilon)$ with respect to $s$, evaluated at $s = 0$ are:

\begin{eqnarray*} 
\frac{\partial g(Cs,\epsilon)}{\partial s} \Bigg|_{s=0} & = & 0 \\ 
\frac{\partial^{2} g(Cs,\epsilon)}{\partial s^{2}} \Bigg|_{s=0} & = & -\frac{1}{4} \frac{\cos{(\epsilon)} C^2}{\sqrt{1-\cos^{2}{(\epsilon)}}} 
\end{eqnarray*}

respectively.

3) Therefore, by considering a Taylor series approximation we have that this error to leading order as $s \to 0$ is:

\begin{eqnarray*}
\int_{\epsilon-\frac{\cot{(\epsilon)} C^2 s^2}{8}+o(s^2)}^\epsilon \Bigg(-\frac{C \cos{(\epsilon)}}{2} \sqrt{\frac{-\cos{(\epsilon)}}{2 \sin{(\epsilon)} (\theta_P-\epsilon)}}  - \cos{(\theta)} \Bigg) s d \theta_P = O(s^2),
 \end{eqnarray*}

where we approximate $g(\frac{C}{t}, \epsilon)$ by $\epsilon-\frac{\cot{(\epsilon)} C^2 s^2}{8}$ and approximate $\frac{C}{2\cos^{-1} (\frac{\cos{(\epsilon)}}{\cos{(\theta_P)}}) t}$ by its leading order term for $|\theta - \epsilon| << 1$ in equation (\ref{eq: theta_exp}).

4) Therefore, substituting $s = \frac{1}{t}$ yields:

\begin{equation} \label{eq:err_ord}
\centering 
\int_{g(\frac{C}{t},\epsilon)}^{\epsilon} \Bigg( \frac{C}{2\cos^{-1} (\frac{\cos{(\epsilon)}}{\cos{( \theta_P)}}) t} - 1 \Bigg) \cos \theta_P \,d\theta_P = O(t^{-2})
\end{equation}

An expression for the leading order behaviour of the difference between $P_{sc}^{u}(t)$ and $P_{sc}^{r}(t)$ is derived as follows:

\begin{eqnarray}
\int_{g(\frac{C}{t},\epsilon)}^{\epsilon} \Bigg(\frac{C}{2\cos^{-1} (\frac{\cos{(\epsilon)}}{\cos{(\theta_P)}}) t} - 1 \Bigg) \cos{(\theta_P)} \,d\theta_P \sim \int_{g(\frac{C}{t},\epsilon)}^{\epsilon} \Bigg[\frac{C}{2 t} \Bigg(-\frac{1}{2} \frac{\cos^{2}{(\epsilon) \sqrt{\frac{2 \sin{(\epsilon)}}{\cos{(\epsilon)}}}}}{\sin{(\epsilon)} \sqrt{\epsilon - \theta_P}} \Bigg) - \cos{( \theta_P)}\Bigg] \,d\theta_P.
\end{eqnarray}

Now substitute $\frac{\cos{(\theta_P)}}{\cos^{-1} (\frac{\cos{(\epsilon)}}{\cos{(\theta_P)}})}$ by the leading order term from its series expansion around $\theta_P = \epsilon$, which is $-\frac{1}{2} \frac{\cos^{2}{(\epsilon)} \sqrt{\frac{-2\sin{(\epsilon)}}{\cos{(\epsilon)}}}}{\sin{(\epsilon)}\sqrt{\theta_P-\epsilon}}$: 

\begin{eqnarray} \label{eq: C_sc}
\int_{g(\frac{C}{t}, \epsilon)}^{\epsilon}\Bigg[\frac{C}{2 t}\Bigg(-\frac{1}{2}\frac{\cos^{2}{(\epsilon) \sqrt{\frac{2\sin{(\epsilon)}}{\cos{(\epsilon)}}}}}{\sin{(\epsilon)}\sqrt{\epsilon - \theta_P}} \Bigg) - \cos{(\theta_P)} \Bigg] \,d\theta_P & \sim & - \frac{C^2 \cos^{2}{(\epsilon)} s^2}{4 \sin{(\epsilon)}} - \frac{C^2 \cos^{2}{(\epsilon)} s^2}{8 \sin{(\epsilon)}} \nonumber \\
& = & -\frac{27 \cos^{2}{(\epsilon)}}{2 \pi^2 t^2 \sin{(\epsilon)}}, 
\end{eqnarray}

where $O(s^2)$ terms are sought, $g(\frac{C}{t}, \epsilon) \sim \epsilon - \frac{1}{8} \frac{\cos{(\epsilon)} C^2}{\sqrt{1-\cos^{2}{(\epsilon)}}} s^{2}$, the value of \(C=\frac{6}{\pi}\) is substituted and \(\sin{\Bigg(\epsilon-\frac{1}{8} \frac{\cos{(\epsilon)}}{\sin{(\epsilon)}} C^2 s^2 \Bigg)} \sim \sin{\epsilon} - \frac{C^2 \cos^{2}{(\epsilon)} s^2}{8 \sin{(\epsilon)}} \) as \( s \to 0 \).

\end{document}